\documentclass[aps,reprint,superscriptaddress,amsmath,amssymb,pra]{revtex4-2}

\usepackage{amsmath}
\usepackage{graphicx}
\usepackage{float}
\usepackage[english]{babel} 

\usepackage[usenames,dvipsnames]{xcolor}
\usepackage[colorlinks=true,citecolor=Cerulean,linkcolor=RubineRed,urlcolor=Cerulean]{hyperref}

\begin{document}

\title{Interacting Bosons on Crystalline and Quasiperiodic Ladders in a Magnetic Field}
\author{Dean Johnstone}
\email{dean.johnstone@polytechnique.edu}
\affiliation{SUPA, Institute of Photonics and Quantum Sciences,
	Heriot-Watt University, Edinburgh, EH14 4AS, United Kingdom}
\affiliation{CPHT, CNRS, École polytechnique, Institut Polytechnique de Paris, 91120 Palaiseau, France}
\author{Patrik \"{O}hberg}
\affiliation{SUPA, Institute of Photonics and Quantum Sciences,
	Heriot-Watt University, Edinburgh, EH14 4AS, United Kingdom}
\author{Callum W. Duncan}
\affiliation{Department of Physics, SUPA and University of Strathclyde, Glasgow G4 0NG, United Kingdom}
\date{\today}

\begin{abstract}
We study a variety of Hofstadter ladders in order to probe the interplay between interactions, an applied magnetic field and crystalline or quasiperiodic geometries. Rotational motion will be induced on charged particles when a magnetic field is present, which can result in exotic distributions of current on a lattice. Typically, the geometry of a ladder lattice is assumed to be homogeneous. In this work, however, we will also study ladders that possess non-uniform bond lengths, in order to study the formation of localised currents. By using Density Matrix Renormalisation Group (DMRG) to characterise the quantum phases, we confirm the presence of the usual vortex and Meissner distributions of current, in which particles circulate within the bulk and around the edge respectively. Furthermore, it is also possible to observe variations to these patterns; which combine both vortex and Meissner order, and the onset of incompressible domains for certain fillings of the lattice. If the bond lengths of a ladder fluctuate, we find substantial differences to the structure of currents. This is a consequence of an inhomogeneous, effective magnetic flux, resulting in preferential localisation of currents throughout the lattice bulk, towards the smaller bond lengths. We then find that incompressible domains can significantly grow in size extent across the parameter space, with currents no longer possessing an extended structure across the longitudinal direction of the ladder.
\end{abstract}

\maketitle

\section{Introduction}
The Hofstadter model is a well-studied problem that describes the motion of charged particles on a 2D lattice when an external magnetic field is present. This model was first studied by Harper and Hofstadter \cite{harper1955,PhysRevB.14.2239}, and has a fractal structure to the energy spectrum versus magnetic flux, in the form of a Hofstadter butterfly. The single-particle properties have been relatively well studied for a variety of lattice geometries, including the presence of flat bands \cite{PhysRevB.54.R17296,PhysRevB.90.075104}, spectral evolution \cite{PhysRevA.95.063628,PhysRevB.98.245145} and the appearance of topological edge states \cite{PhysRevLett.71.3697,PhysRevLett.124.036803,PhysRevLett.123.196401}. The extension of the Hofstadter model to include interactions between particles is an interesting problem to consider, as it can allow for the formation of quantum phases with exotic distributions of current, i.e. phases with a preferential flow of tunnelling. For bosons, this was first studied in the context of Josephson junction arrays \cite{PhysRevB.42.4797,PhysRevB.52.10433,PhysRevLett.75.3930,PhysRevB.33.3125,PhysRevB.64.144515}. More recently, there has been interest in the use of ultracold gases to emulate these properties. By trapping bosonic atoms in an optical lattice, it has been possible to realise highly controllable experimental frameworks that simulate Bose-Hubbard models \cite{Jaksch1998,greiner2002}. However, due to atoms being charge neutral, magnetic fields must be introduced by artificial means. This has included the use of rotating Bose-Einstein condensates \cite{PhysRevLett.84.806,Shaeer2001,PhysRevLett.92.040404,PhysRevLett.92.050403}, laser assisted processes \cite{goldman2014light,PhysRevA.73.025602,PhysRevLett.102.130401,jaksch2003creation}, driven systems \cite{PhysRevLett.108.225304,PhysRevLett.109.145301,PhysRevX.4.031027,creffield2016realization} and internal degrees of freedom \cite{Stuhl2015,Mancini2015,PhysRevLett.117.220401,PhysRevLett.115.195303}. By employing these techniques, it has been possible to realise interacting Hofstadter models with bosons \cite{PhysRevLett.107.255301,PhysRevLett.108.225303,PhysRevLett.111.185301,PhysRevLett.111.185302}, usually for square or periodic lattice geometries.

The consideration of ladder geometries represents a toy model to probe the interplay between interactions and an applied magnetic field. Indeed, the first experimental observation in ultracold atoms of chiral currents was reported by Atala \textit{et al} in 2014 \cite{atala2014observation} for a ladder. Further works since then have studied the numerical and theoretical properties in detail using DMRG \cite{PhysRevB.91.140406,PhysRevB.92.060506,Orignac_2016,PhysRevB.96.014518,PhysRevA.94.063628,PhysRevLett.115.190402} and bosonisation techniques \cite{PhysRevLett.111.150601,PhysRevB.91.054520,tokuno2014ground,PhysRevA.89.063617,PhysRevA.93.053629}, with particular interest in the underlying distributions of current for different quantum phases. In relation to the currents, topological invariants have also been measured in ladder systems, notably the many-body Chern number \cite{aidelsburger2015measuring,genkina2019imaging,chalopin2020probing,SciPostPhys.3.2.012}. For the aforementioned studies, the geometry of the ladder itself is assumed to be uniform, with no underlying spatial dependence of the model parameters. For this work, we will consider a range of ladder geometries which are not uniform, including those that are quasiperiodic. By varying the horizontal bond lengths of a ladder, it is possible to introduce another form of preferential localisation into the system. In particular, within a single-particle picture, it is known that non-uniform, quasiperiodic lattices can host in-gap, topological states that are localised in the bulk, rather than the edge. \cite{PhysRevB.106.045149,duncan2020topological}. It is natural to expect other exotic properties to arise on non-uniform ladders when interactions are present.

Here, we present our results as follows. In Sec.~\ref{sc_ladders}, we introduce the different kinds of Hofstadter ladder geometries for interacting bosons and discuss how ground states can be found via DMRG. Following on from this, we then outline several order parameters that can be used to characterise different quantum phases and current distributions in Sec.~\ref{sc_phases}. We then show our first set of results in Sec.~\ref{sc_currents}, which focuses on the behaviour and locality of currents for the ladder systems, including momentum profiles. Finally, phase diagrams are presented in Sec.~\ref{sc_mDiag} over a larger range of parameters, before we end with our conclusions in Sec.~\ref{sc_conc}.

\section{Ladder Systems} \label{sc_ladders}

\subsection{Interacting Hofstadter Model}
\begin{figure}[h!]
	\centering
	\makebox[0pt]{\includegraphics[width=0.9\linewidth]{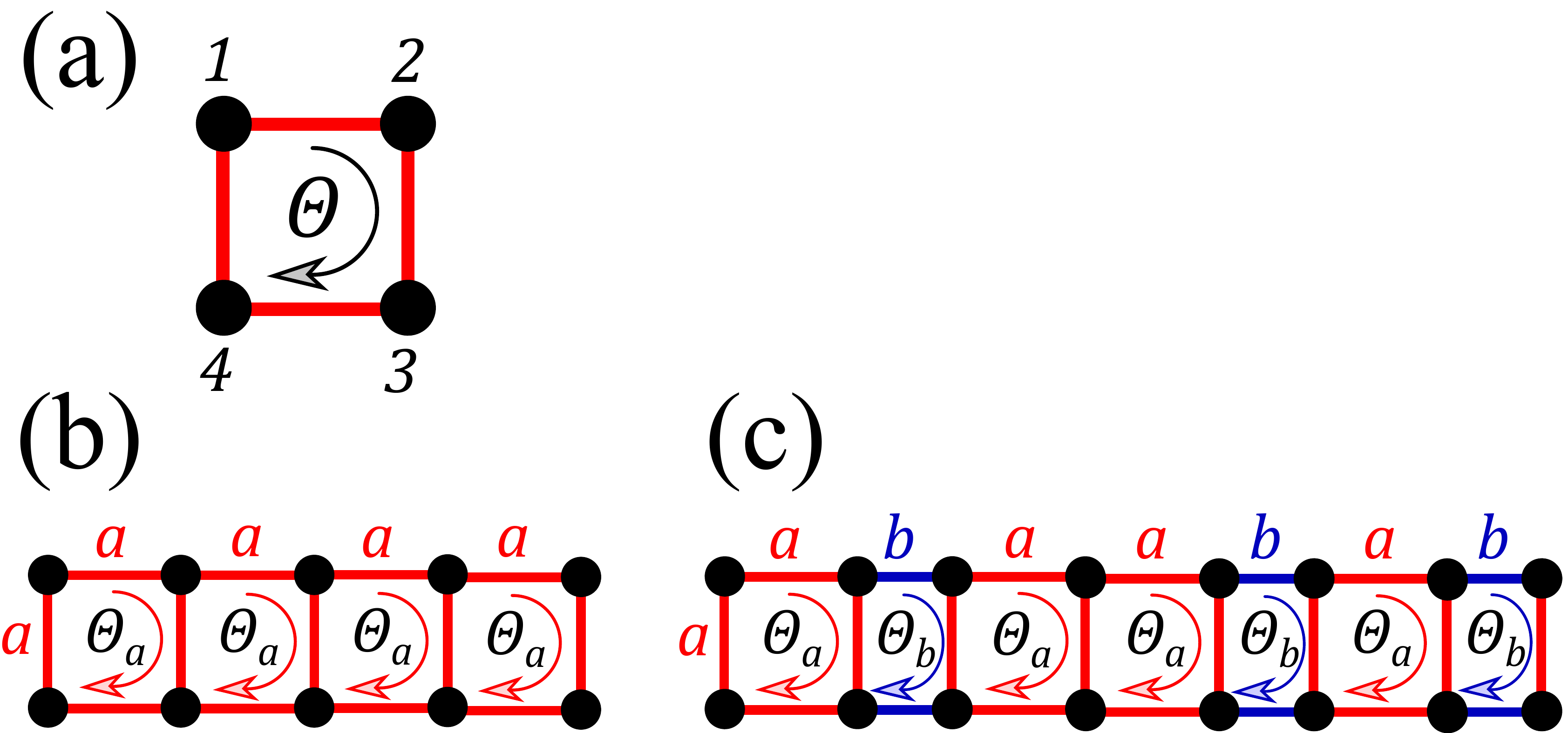}}
	\caption{Illustration of the encircling phases for different kinds of ladder systems, with bond lengths $a$ (red) and $b$ (blue). If a particle circulates the square tile in (a), it will acquire a unique phase factor $\Theta$. We consider (b) uniform and (c) non-uniform distributions of $a$ and $b$, which will produce the encircling phases $\Theta_a$ and $\Theta_b$ on the ladder due to the geometrical dependence of the Peierls phases in Eq.~\eqref{eq_pPhases}.}
	\label{fig_model_ex}
\end{figure}

For a lattice of size $L = L_x \times L_y$ in the presence of a magnetic field, the system will be described by an interacting Hofstadter model of the form
\begin{equation}	\label{eq_hm}
\mathcal{\hat{H}} = -\sum_{\langle i,j \rangle}^L J_{ij} e^{i \theta_{ij}} \hat{b}^\dagger_i \hat{b}_j - \mu \sum^L_i \hat{n}_i + \frac{U}{2}\sum^L_i \hat{n}_i (\hat{n}_i- 1),
\end{equation}
where $U$ is the on-site energy, $J_{ij}$ are the tunnelling coefficients, $\mu$ is the chemical potential, $\hat{b}_{i} \, (\hat{b}_{i}^\dagger)$ are the bosonic atom destruction (creation) operators at site $i$, $\langle i,j \rangle$ denotes nearest-neighbour summations across lattice bonds and $\theta_{ij}$ are the Peierls phase factors \cite{Peierls1933}. The Peierls phases depend on the magnetic vector potential $\mathbf{A}(\mathbf{r})$, with $\mathbf{r}$ being a spatial coordinate. Their form can be written as
\begin{equation}
\theta_{ij} = \int^{\textbf{r}_i}_{\textbf{r}_j} \mathbf{A}(\mathbf{r}) \cdot d\mathbf{r},
\end{equation}
where $\textbf{r}_i$ are coordinates of a lattice site $i$. By working in a Landau gauge of
\begin{equation}
\mathbf{A}(\mathbf{r}) = (0, \, Bx,\, 0),
\end{equation}
with $B$ denoting the magnetic field strength and $x$ being the $x$-coordinate, the Peierls phases can be expressed as
\begin{equation} \label{eq_pPhases}
\theta_{ij} = \dfrac{\phi}{2 A} (x_i + x_j)(y_j - y_i),
\end{equation}
where $(x_i, \, y_i)$ are the $x$/$y$ spatial coordinates at site $i$ and we have introduced the magnetic flux $\phi = BA$, for a reference area $A$ that corresponds to a square tile of the lattice, see Fig.~\ref{fig_model_ex}(a). Throughout this work, we set $\hbar = q = 1$ and assume that the magnetic flux is measured in units of the flux quantum $\phi_0 = 2\pi$.

We will consider ladder geometries with $L_y=2$, as depicted in Fig.~\ref{fig_model_ex}, with two distinct horizontal bond lengths $a$ and $b$. If a particle tunnels across the edges of a single tile, it will acquire an encircling phase $\Theta = \theta_{1,2} + \theta_{2,3} - \theta_{3,4} - \theta_{4,1}$, labelled according to the vertices in Fig.~\ref{fig_model_ex}(a). Due to the geometrical dependence of the Peierls phases $\theta_{ij}$, different horizontal bond lengths will produce distinct encircling phases. Non-uniform ladders have been studied for the case of sign-staggered distributions of $\Theta$. The staggered arrangement of $\Theta$ can be realised with ultracold atom experiments that use time-dependent fluctuations of an external potential \cite{PhysRevA.81.023404}. The effective magnetic field is then zero, and chiral phases have been reported \cite{PhysRevA.85.041602,PhysRevB.87.174501,PhysRevB.91.054520,PhysRevA.81.023404,PhysRevA.82.063625}, in which the loop currents possess interesting, staggered structures that break time-reversal symmetry.

We will consider different distributions of the bond lengths $a$ and $b$, in accordance to a binary function $X(m)$ for the $m$th bond. This binary sequence allows for a mapping to the horizontal bond lengths as $0 \rightarrow a$ and $1 \rightarrow b$, with an example lattice in Fig.~\ref{fig_model_ex}(c). We will consider two distributions of $X(m)$. The first is that of a superlattice, which has the two bond lengths oscillating in a periodic manner
\begin{equation} \label{eq_x_sp}
\begin{aligned}
X_s(m) = m - 1 - 2 \lfloor \frac{m-1}{2} \rfloor \\ = [0, \, 1, \, 0, \, 1, \, 0, \, 1, \, 0, \, 1, \, \dots],
\end{aligned}
\end{equation}
where $\lfloor ... \rfloor$ is the floor operation. We also consider a quasiperiodic distribution of bond lengths, defined by the Fibonacci word
\begin{equation} \label{eq_x_qc}
\begin{aligned}
X_q(m) = 2 + \lfloor m \tau \rfloor - \lfloor (m+1) \tau \rfloor \\ = [0, \, 1, \, 0, \, 0, \, 1, \, 0, \, 1, \, 0, \, 0, \, 1, \, \dots],
\end{aligned}
\end{equation}
where $\tau=\frac{1+\sqrt{5}}{2}$ is the golden ratio. The quasiperiodic ladder will break translational invariance, but will retain long-range order, analogous to 2D quasicrystalline tilings \cite{DEBRUIJN198139,DEBRUIJN198153,penrose1974role,Robinson1971}.

We will also incorporate a qualitative, geometrical scaling of the tunnelling coefficients $J_{ij}$ as 
\begin{equation}
J_{ij} = \frac{J}{| \mathbf{r}_i - \mathbf{r}_j  |}.
\end{equation}
This emulates the spatial dependence of tunnelling coefficients in optical lattice experiments, where it is usual for $J_{ij}$ to scale with the relative separation between sites \cite{PhysRev.52.191}. Note, however, that similar results can be observed without this requirement. Finally, throughout this work, we also use the convention that $a=1$ produces the tunnelling rate $J$, with vertical bond lengths also kept constant at $a$. The only fluctuations that then arise are the $b$ coefficients for $a/b \neq 1$.

\begin{figure}[h!]
	\centering
	\makebox[0pt]{\includegraphics[width=0.99\linewidth]{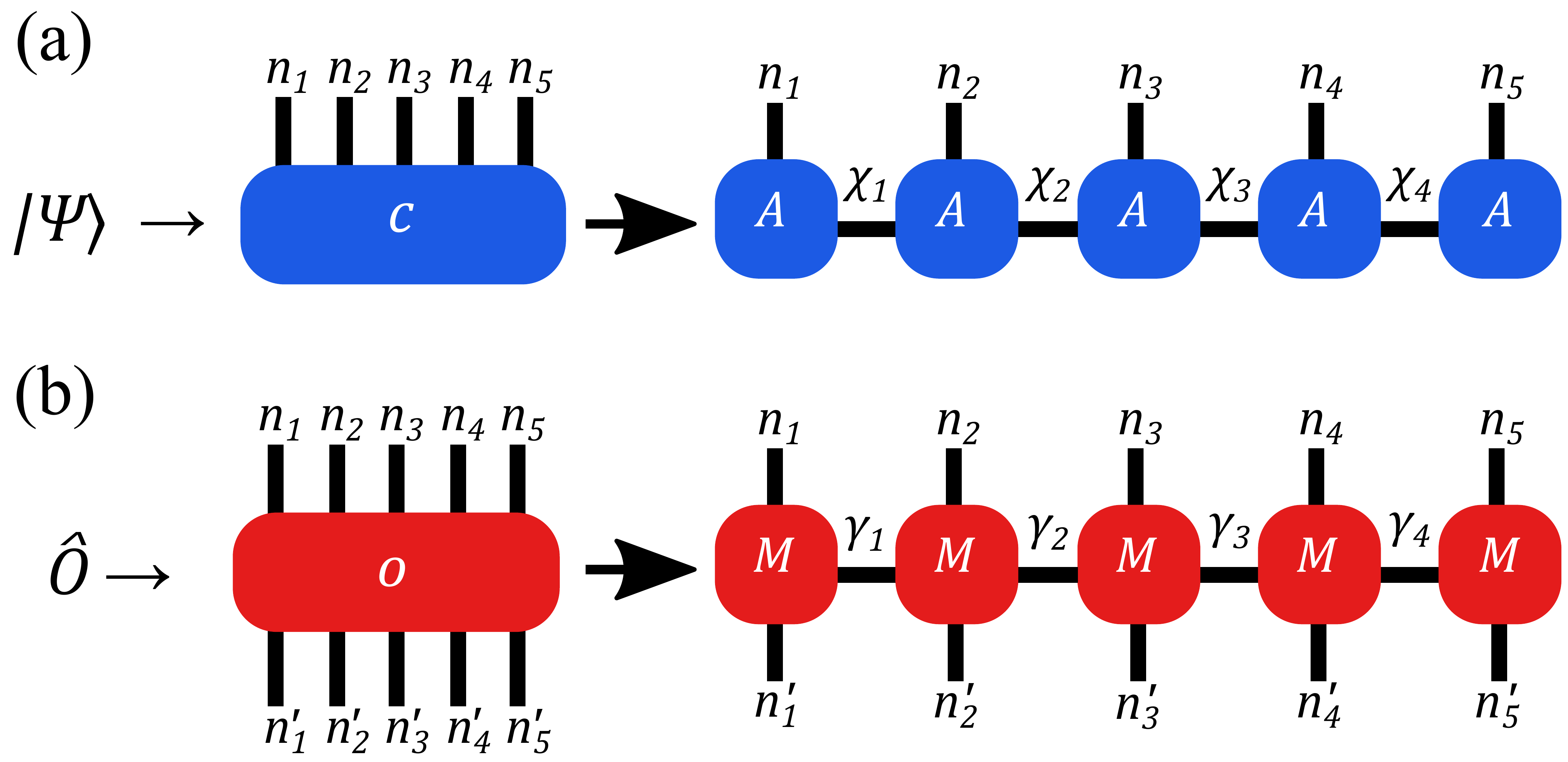}}
	\caption{Depiction of (a) a MPS and (b) MPO tensor network for a $L=5$ 1D lattice. For each case, the many-body wavefunction $| \Psi \rangle$ and operator $\hat{O}$ are expressed in a graphical form, with legs representing different indices of the tensors $c$ and $o$. In the tensor network, $c$ and $o$ are rewritten as a contraction of individual tensors $A$ or $M$ at a lattice site, where vertical legs $n_i$ and $n'_i$ denote the physical dimensions at a certain site, i.e. the local Hilbert space. Finally, horizontal legs $\chi_i$ and $\gamma_i$ are the bond dimensions, which capture the entanglement and correlations across the tensor network.}
	\label{fig_dmrgMps}
\end{figure}

\subsection{Density Matrix Renormalisation Group}
Ground states of interacting models in the presence of magnetic fields can possess exotic structures and localisation across the lattice. To probe these properties, we will calculate ground states numerically using the well-known framework of DMRG, which can overcome the exponential scaling of the Hilbert space in certain scenarios \cite{PhysRevLett.69.2863,PhysRevB.48.10345}. Specifically, we will adapt the Matrix Product State (MPS) based ansatz for the many-body wavefunction $| \Psi \rangle$, which allows for a wavefunction coefficient to be expressed as a series of contracted tensors \cite{PhysRevLett.75.3537,Dukelsky_1998}, as per Fig.~\ref{fig_dmrgMps}(a). Likewise, Hamiltonians or generic operators $\hat{O}$ can also be expressed in a similar form, known as Matrix Product Operators (MPOs) \cite{PhysRevLett.93.207205,PhysRevLett.93.207204}, shown in Fig.~\ref{fig_dmrgMps}(b). In both cases, the vertical bonds of the local MPS/MPO tensors represent the local Hilbert space of a site in the Fock number basis. The horizontal bonds are known as the bond dimensions, and are used as a variational parameter to control the degree of correlations/entanglement retained within the calculation. In practice, the bond dimensions can be significantly compressed and still represent near-exact ground states \cite{Verstraete2008,RevModPhys.77.259} or operators \cite{PhysRevB.95.035129}.

\begin{figure}[h!]
	\centering
	\makebox[0pt]{\includegraphics[width=0.99\linewidth]{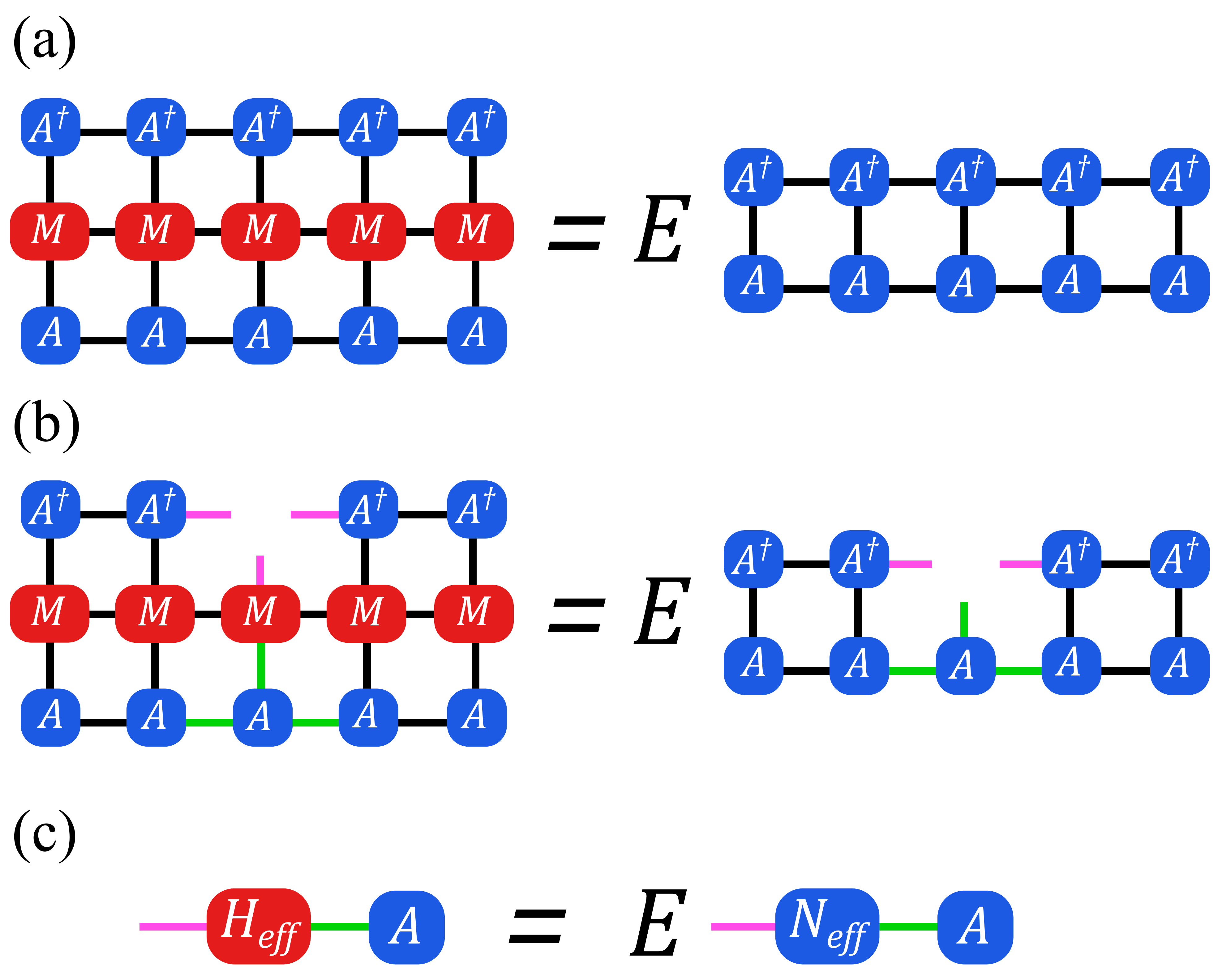}}
	\caption{Solving for the ground state, given a MPS wavefunction. We first write Eq.~\eqref{eq_dmrgOptim} as a tensor network in (a), and remove a single $A^\dagger$ tensor from the system, which is shown in (b) for site $3$. By then contracting tensors (black bonds) and fusing together the purple and green bonds, the problem can be reduced to a generalised eigenvalue problem $H_{eff} A = E N_{eff} A$. In practice, gauge transforms of the MPS can also transform $N_{eff} \rightarrow I$, where $I$ is the identity matrix, which instead produces a standard eigenvalue problem.}
	\label{fig_dmrgOptim}
\end{figure}

To solve for the ground state, we are then interested in minimising the Schr\"{o}dinger equation
\begin{equation} \label{eq_dmrgOptim}
\langle \Psi | \mathcal{\hat{H}} | \Psi \rangle = E \langle \Psi | \Psi \rangle,
\end{equation}
which is depicted in Fig.~\ref{fig_dmrgOptim}(a) as a tensor network, where $E$ is the eigenenergy. For practical purposes, a single tensor is usually removed from one of the MPS networks in Fig.~\ref{fig_dmrgOptim}(b), allowing for the minimisation procedure to be reduced to an eigenvalue problem for a local site, as per Fig.~\ref{fig_dmrgOptim}(c). Given an initial wavefunction, the ground state can then be found by optimising each local MPS tensor for each site until the energy converges.

\begin{figure}[t!]
	\centering
	\makebox[0pt]{\includegraphics[width=0.8\linewidth]{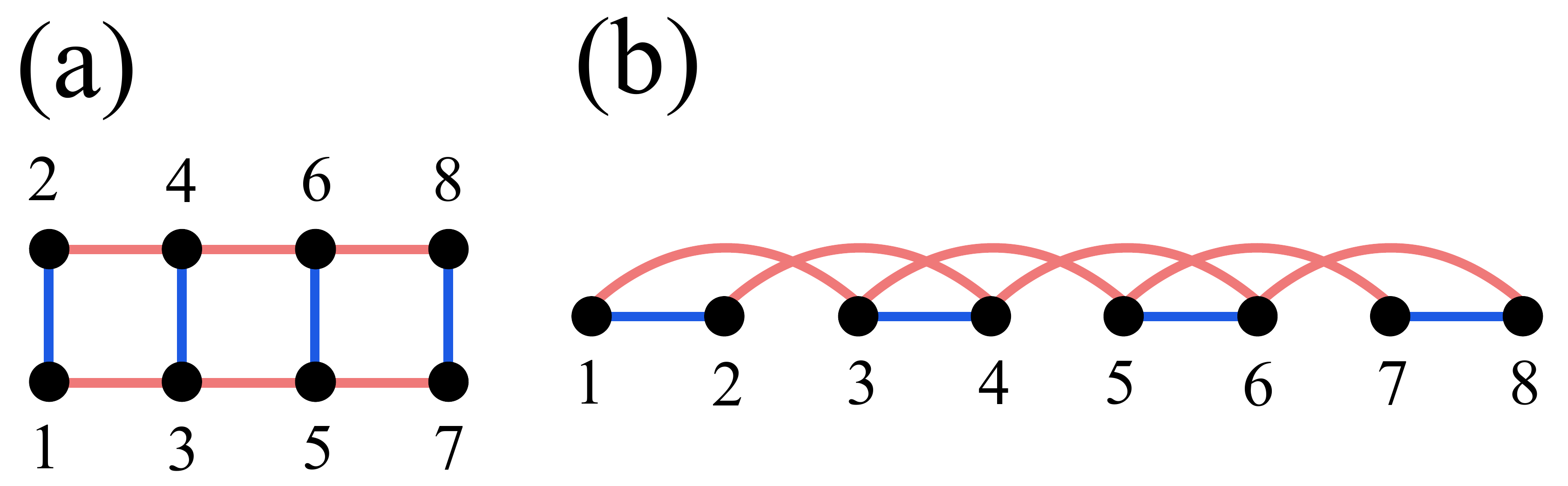}}
	\caption{1D mapping of lattice sites and bonds for a (a) $L = 2 \times 4$ square lattice. Each site has a unique index between $1$ and $L$, which are mapped to the effective 1D system in (b). Red and blue connections between sites denote the horizontal ($x$-direction) and vertical ($y$-direction) bonds respectively.}
	\label{fig_1dMap}
\end{figure}

Typically, DMRG algorithms are applied to 1D lattice models. However, their extension to higher dimensions is also possible, provided an effective 1D mapping does not generate exceedingly long-range terms. We illustrate this idea in Fig.~\ref{fig_1dMap} for a ladder system. In 1D, the ladder system then has nearest-neighbour tunnelling for processes in the $y$-direction and next-nearest-neighbour tunnelling for processes in the $x$-direction.

\section{Observables and Phases} \label{sc_phases}
Given a MPS ground state, it is possible to probe a range of observables that can characterise the structure and localisation of quantum phases. First, the correlation function of a state can be calculated as
\begin{equation} \label{eq_c}
c_{ij} = \langle \hat{b}^\dagger_i \hat{b}_j \rangle,
\end{equation}
which can reveal the presence of quantum correlations. The correlation function is closely related to the density profile of a state in momentum space, which has the form
\begin{equation} \label{eq_nk}
n(\mathbf{k}) = \frac{1}{L} \sum_{ij} e^{i \mathbf{k} \cdot (\mathbf{r}_i - \mathbf{r}_j ) } \langle \hat{b}^\dagger_i \hat{b}_j \rangle,
\end{equation}
where $\mathbf{k}$ is the momentum. The momentum profile is especially relevant for experimental protocols with ultracold atoms, as it can be directly measured from time-of-flight absorption images \cite{greiner2002,PhysRevLett.100.120402,folling2005spatial}. These procedures can be used to identify whether or not a given phase is a Mott-Insulator (MI) or Superfluid (SF). \cite{Jaksch1998,greiner2002}. A SF possesses macroscopic phase coherence and long-range correlations, resulting in a momentum profile with a sharp peak at $|\mathbf{k}|=0$. The compressibility
\begin{equation} \label{eq_comp}
\kappa = \frac{1}{L}\frac{\partial \langle \hat{N} \rangle}{\partial \mu},
\end{equation}
will be finite, with $\langle \hat{N} \rangle$ the total particle number. The MI contains no correlations, and instead has a flat momentum profile due to particle immobility. As a result, $\kappa$ is zero and the phase will be incompressible. The interacting Hofstadter model is known to host different kinds of incompressible phases from the usual MI. This can include Charge Density Waves \cite{PhysRevA.94.063628,PhysRevLett.119.073401,PhysRevA.90.053623}, incompressible Meissner or vortex domains \cite{PhysRevB.91.140406,PhysRevLett.111.150601} and topologically non-trivial Chern insulators or quantum Hall states \cite{PhysRevA.90.053623,PhysRevB.91.054520,PhysRevX.7.021033,PhysRevB.96.014524,PhysRevLett.94.086803}. While these phases are incompressible, they will typically contain finite currents and non-integer on-site densities. Due to this support of inherent transport, we will generally refer to these phases as incompressible, rather than insulating, i.e. the transport/currents are robust against particle number fluctuations.

\subsection{Current Distributions}
Charged particles in a magnetic field will flow in a cyclical manner, meaning that there is a preferential flow of tunnelling, or current present in the system. This can be quantified from Heisenbergs equation of motion for the on-site densities
\begin{equation}
\frac{d \hat{n}_i}{dt} = i[\mathcal{\hat{H}},\hat{n}_i] \equiv \sum_j \hat{j}_{ij},
\end{equation}
where the sites ($i$, $j$) are connected by bonds and the current operator $\hat{j}_{ij}$ between sites is defined as
\begin{equation}
\hat{j}_{ij} = i J_{ij} e^{i \theta_{ij}} \hat{b}^\dagger_i \hat{b}_j - H.c.
\end{equation}
The expectation values of $\hat{j}_{ij}$ denote local bond currents in the system, and will be the key order parameters in classifying different patterns of current that appear within MI or SF phases. From these, we can first define an average absolute current across vertical bonds as 
\begin{equation}
j_{V} = \frac{1}{L_x} \sum_{x_i - x_j = 0} |j_{ij}|,
\end{equation}
where $j_{ij} = \frac{\langle \hat{j}_{ij} \rangle}{J_{ij}}$ and the summation only accounts for lattice sites connected across the $y$ direction, i.e. if the $x_i=x_j$. In a similar manner, average row currents (denoted by the subscript $r$ or $R$) across the $x$ direction can be defined as
\begin{equation}
j_{r}^m = \frac{1}{L_x-1} \sum_{x_i = (m-1) L_y, \, y_i - y_j = 0} j_{ij}
\end{equation}
and
\begin{equation}
j_{R}^m = \frac{1}{L_x-1} \sum_{x_i = (m-1) L_y, \, y_i - y_j = 0} | j_{ij} |,
\end{equation}
where $j_{r}^m$ is the average current across row $m$ that takes into account the sign of each $j_{ij}$, and $j_{R}^m$ is the average absolute current across row $m$ that takes the absolute value of each $j_{ij}$. The summations for $j_{r}^m$ and $j_{R}^m$ only account for sites with the same $y$-coordinate, for fixed $x_i=(m-1)L_y$, i.e. a row of the ladder. From this, the average absolute current across all horizontal bonds can then be written as
\begin{equation}
j_{H} = \frac{1}{L_y} \sum_{m}^{L_y} j_{R}^m.
\end{equation}
Furthermore, we can also characterise a Meissner, or chiral current in the system as
\begin{equation}
j_{M} = \frac{1}{2 L_y} (j_{r}^1 - j_{r}^2),
\end{equation}
which determines the current encircling the two horizontal rows of the ladder. Note, the extra prefactor of $2$ in the denominator is introduced to ensure that $|j_{M}| \leq j_{H}$, which will be important for the discussion of current distributions later in this section.

To better characterise the localisation of bond currents across the ladder, we will also look at current locality functions. The first measure we use determines the fraction of blocked current channels that are present on the ladder, i.e. if a lattice site has at least $1$ blocked path of current. From this, we define $j_L$, which is normalised between $1$ (no currents) and $0$ (all bonds have finite currents). To calculate $j_L$, we iterate through each lattice site and count the number of finite ingoing/outgoing currents from the bonds. If the number of finite ingoing/outgoing currents of a site is greater than $1$, we assign a value of $0$ to a discrete function $S_i$ for site $i$, otherwise we assign a value of $1$ to $S_i$ if there are no ingoing/outgoing currents. $j_L$ is then taken as the average value of $S_i$.

Next, we also calculate the locality of finite currents across the vertical rungs of the ladder. This measure, denoted as $j_E$ is calculated by taking the average position of vertical currents, where the coordinates are normalised between $0$ (centre of ladder) and $1$ (edges of ladder). If a phase has finite currents across all bonds, $j_L=1$ and 
\begin{equation}
j_{E} = \frac{2}{L_x \lfloor L_x/2 \rfloor} \sum_{n=0}^{\lfloor L_x/2 \rfloor} n \approx 0.5.
\end{equation}
On the other hand, if currents are absent across certain bonds, both $j_{E}$ and $j_{L}$ can take unique values. For currents with edge localisation across the $x$-direction, both $j_{E}$ and $j_{L}$ will converge to $1$. For other kinds of localised currents, $j_{E} \neq 1$ and $j_{L} > 0$. To better illustrate these ideas, we will now turn our attention to the current distributions that may arise in the interacting Hofstadter model.

\begin{figure}[t!]
	\centering
	\makebox[0pt]{\includegraphics[width=0.9\linewidth]{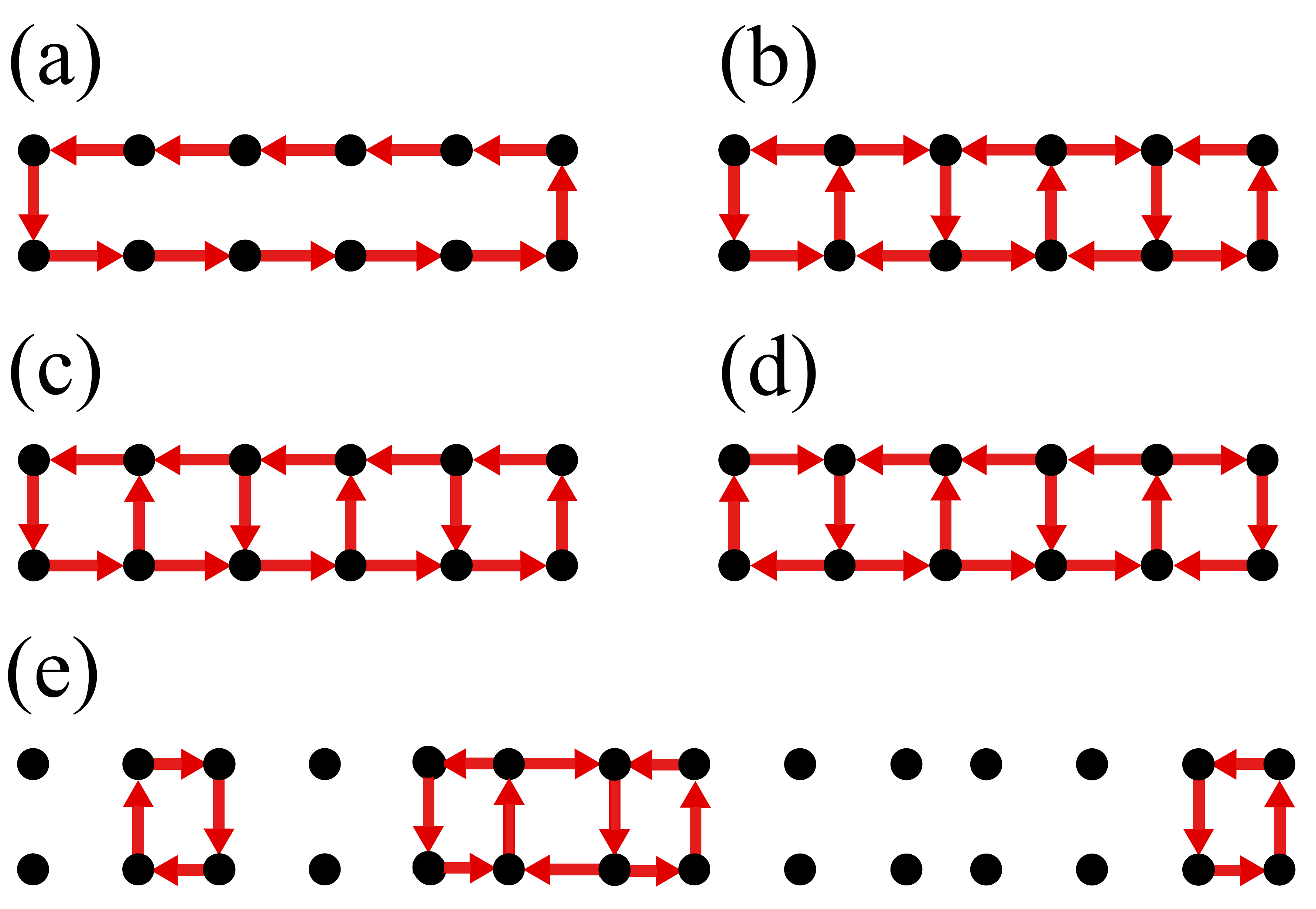}}
	\caption{Illustration of current distributions in the interacting Hofstadter model for a (a-d) periodic ladder of size $L = 2 \times 6$ and (e) a quasiperiodic ladder with $L = 2 \times 14$. Red bonds indicate finite currents $j_{ij}$ across the ladder, with arrows denoting paths of positive current. Note, the magnitude of currents across the ladder may not necessarily be equal. We show (a) Meissner, (b) Vortex, (c) strong Vortex-Meissner, (d) weak Vortex-Meissner and (e) localised current patterns. The localised currents in (e) can contain vortex or vortex-Meissner distributions of current.}
	\label{fig_currentDists}
\end{figure}

Interacting lattice models in the presence of magnetic fields will give rise to exotic distributions of current, in conjunction with the usual MI and SF phases of Bose-Hubbard models. The two primary current distributions are that of the bosonic Meissner and vortex arrangements \cite{PhysRevB.64.144515}, which are based on the Meissner effect \cite{meissner1933neuer,PhysRev.108.1175} and Abrikosov vortices \cite{rjabinin1935magnetic,ginzburg2009theory,RevModPhys.76.975} for type-I and type-II superconductors. In a Meissner domain, currents are localised towards the edges of a ladder, as shown in Fig.~\ref{fig_currentDists}(a). Currents across vertical bonds will effectively be zero, apart from the left/right edges, i.e. $j_{V} \approx 0$ and both $j_{E}$ and $j_L$ are $\approx 1$. Furthermore, currents across horizontal bonds lie in equal but opposite directions for each row of the ladder, which produces a finite Meissner current $j_{M}$.

\begin{table}[h!]
	\centering
	\begin{tabular}{||c c c c c c||} 
		\hline
		Type & $|j_M|$ & $j_V$ & $j_H$ & $j_E$ & $j_L$ \\ [0.5ex] 
		\hline\hline
		Vortex (V) & $= 0$ & $> 0$ & $> 0$ & $= 0.5$ & $= 0$ \\ 
		Meissner (M) & $> 0$ & $= 0$ & $= |j_M|$ & $= 1$ & $= 1$ \\
		strong Vortex-Meissner (sVM) & $> 0$ & $> 0$ & $= |j_M|$ & $= 0.5$ & $= 0$ \\
		weak Vortex-Meissner (wVM) & $> 0$ & $> 0$ & $< |j_M|$ & $= 0.5$ & $= 0$ \\
		Trivial (T) & $= 0$ & $= 0$ & $= 0$ & $= 0.5$ & $= 1$ \\ [1ex] 
		\hline
	\end{tabular}
	\caption{Different distributions of current that can arise in the interacting Hofstadter model.}
	\label{table_phases}
\end{table}

The vortex pattern will have a similar structure to the one in Fig.~\ref{fig_currentDists}(b). Here, all horizontal currents will be anti-aligned, leading to a zero Meissner current. Vertical currents are also finite, leading to a finite value of $j_{V}$. For certain parameter regimes of $\mu/U$ and $\phi / \phi_0$, we also find several variations of the vortex pattern that incorporates a degree of Meissner ordering, with examples in Figs.~\ref{fig_currentDists}(c,d). These are referred to as vortex-Meissner patterns with strong or weak order, depending on the relative scaling of $|j_M|$ to $j_H$. Note, weak or strong vortex-Meissner domains are not distinct phases, but instead measure crossover domains within a vortex-Meissner pattern. For $|j_M| = j_H$, all horizontal rows of current are aligned in opposite directions, and the Meissner order is strong. Otherwise, if $|j_M| \le j_H$, there will be an anti-alignment of certain horizontal currrents, i.e. a weaker Meissner order. In the limiting case of $|j_M| = 0$, we simply have a vortex pattern. Finally, if no finite currents are present in the system, we simply have a standard SF or MI phase, which is trivial from the point of view of the current. In Table~\ref{table_phases}, we summarise the different distributions of current that have been outlined. Note, non-uniform bond lengths may also allow for the formation of localised currents, with an example in Fig.~\ref{fig_currentDists}(e). The current distributions of localised domains may be that of either vortex or vortex-Meissner patterns, with $j_L > 0$ and $j_E \neq 1$. Furthermore, if $\kappa$ is finite, then the phase will resemble that of a Bose-Glass (BG); i.e. localised, circulating SF domains and the absence of macroscopic phase coherence \cite{PhysRevB.55.R11981,PhysRevB.40.546}.
		


\begin{figure}[t]
	\centering
	\makebox[0pt]{\includegraphics[width=0.99\linewidth]{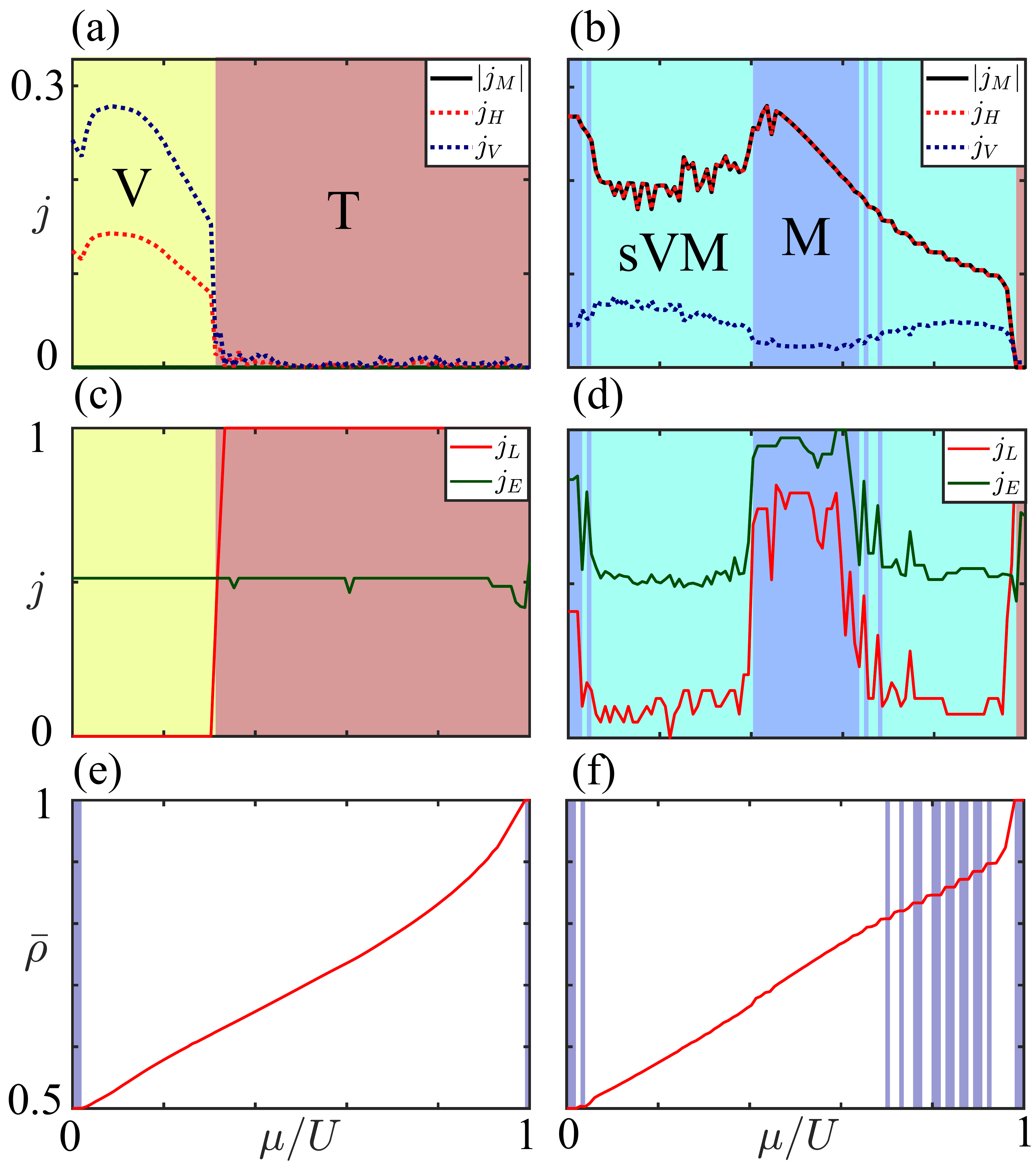}}
	\caption{Plots of (a,b) current order parameters, (c,d) locality measures and (e,f) the average density over a range of $\mu/U$, with fixed (a,c,e) $J/U = 0.44$ and $\phi/\phi_0=0.5$, and (b,d,f) $J/U=0.4$ and $\phi/\phi_0=0.25$. Current distributions are coloured according to the definitions in Table~\ref{table_phases}, where $\mu/U$ variations can change the underlying pattern of current. Incompressible domains are also denoted by shaded areas in (e,f), i.e. the plateaus of $\bar{\rho}$.}
	\label{figure_phi_H}
\end{figure}

\section{Behaviour of Currents} \label{sc_currents}
We begin our results by considering examples of ground state phases on ladders and their scaling over a smaller range of parameters, for hard-core bosons. We will fix $L_x \approx 40$ sites, where fluctuations around this number will be introduced to keep the system commensurate with the magnetic unit cell for superlattice and homogeneous distributions of $X(m)$. While fixed system sizes will be considered in this work, we note that equivalent results have been observed for larger $L_x$.

\begin{figure}[t]
	\centering
	\makebox[0pt]{\includegraphics[width=0.75\linewidth]{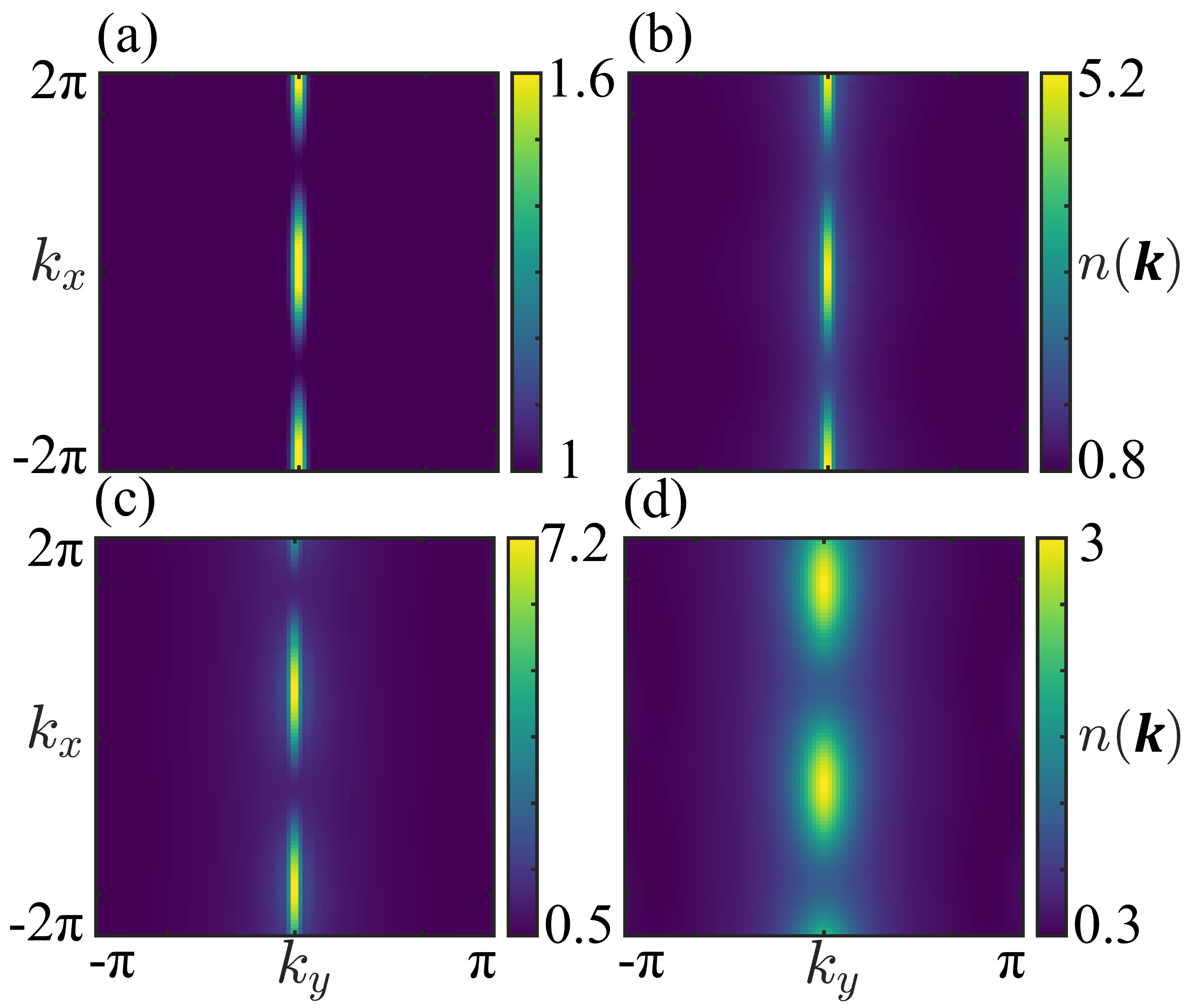}}
	\caption{Momentum profiles $n(\mathbf{k})$ for different phases when $\phi/\phi_0=0.5$ and $J/U=0.44$, including the (a) MI phase at $\mu/U=0.98$ with trivial currents, (b) SF phase at $\mu/U=0.6$ with trivial currents, (c) SF phase at $\mu/U=0.15$ with vortex currents and (d) incompressible phase at $\mu/U=0$ with vortex currents. If the phase is incompressible, $n(\mathbf{k})$ will be relatively flat and delocalised. On the other hand, for compressible, or SF-like phases, the peak in $n(\mathbf{k})$ will become very thin and localised. When finite currents are present, the momentum peak is shifted away $\mathbf{k}=(0,0)$ due to chiral motion.}
	\label{figure_nk1_H}
\end{figure}

\subsection{Homogeneous Ladder ($a/b=1$)}
We will briefly cover some results for the well-studied case of homogeneous $X(m)$, where $a/b=1$. In Fig.~\ref{figure_phi_H}, the current order parameters are plotted for two distinct fluxes $\phi/\phi_0=0.5$ and $\phi/\phi_0=0.25$, with current patterns annotated according to Table~\ref{table_phases}. For Fig.~\ref{figure_phi_H}(a), we have $J/U=0.44$ and $\phi/\phi_0=0.5$, leading to the stabilisation of vortex currents for a range of $\mu/U$, which have zero $j_M$. All currents will be extended across the lattice, as shown with Fig.~\ref{figure_phi_H}(c) for $j_L=0$ and $j_E\approx 0.5$. By increasing $\mu/U$ beyond $0.35$, the currents will decay and oscillate around zero, which marks the onset of a trivial pattern. In Fig.~\ref{figure_phi_H}(e), we also plot the average density $\bar{\rho}$, which varies continuously across a large range of $\mu/U$, showing that the underlying phase is that of a SF. Near $\mu/U=0$ and $\mu/U=1$, $\bar{\rho}$ will begin to plateau, implying that $\kappa \rightarrow 0$, i.e. the onset of incompressible phases. To better understand the differences between the phases and current patterns, we also plot momentum profiles $n(\mathbf{k})$ in Fig.~\ref{figure_nk1_H} at different $\mu/U$. For Fig.~\ref{figure_nk1_H}(a) at $\mu/U=0.98$, the momentum profile is relatively flat and centred at $\mathbf{k}=(0,0)$, indicating a MI phase. When $\mu/U$ is decreased to $0.6$, the system enters a SF region in Fig.~\ref{figure_nk1_H}(b), and $n(\mathbf{k})$ becomes localised to $\mathbf{k}=(0,0)$, denoting the onset of phase coherence. By further tuning $\mu/U=0.15$ into the vortex region, $n(\mathbf{k})$ in Fig.~\ref{figure_nk1_H}(c) will be shifted to $\mathbf{k}\approx(\pi/2,0)$ due to the presence of chiral currents. As we enter the other incompressible domain in Fig.~\ref{figure_nk1_H}(d) at $\mu/U=0$, we observe similar properties, but with a shifted momentum peak to $\mathbf{k}\approx(-5\pi/8, 0)$ due to a sign-flipping of currents. Finally, this $n(\mathbf{k})$ is also more extended when compared against SF phases, analogous to the MI from Fig.~\ref{figure_nk1_H}(a).

\begin{figure}[t]
	\centering
	\makebox[0pt]{\includegraphics[width=0.99\linewidth]{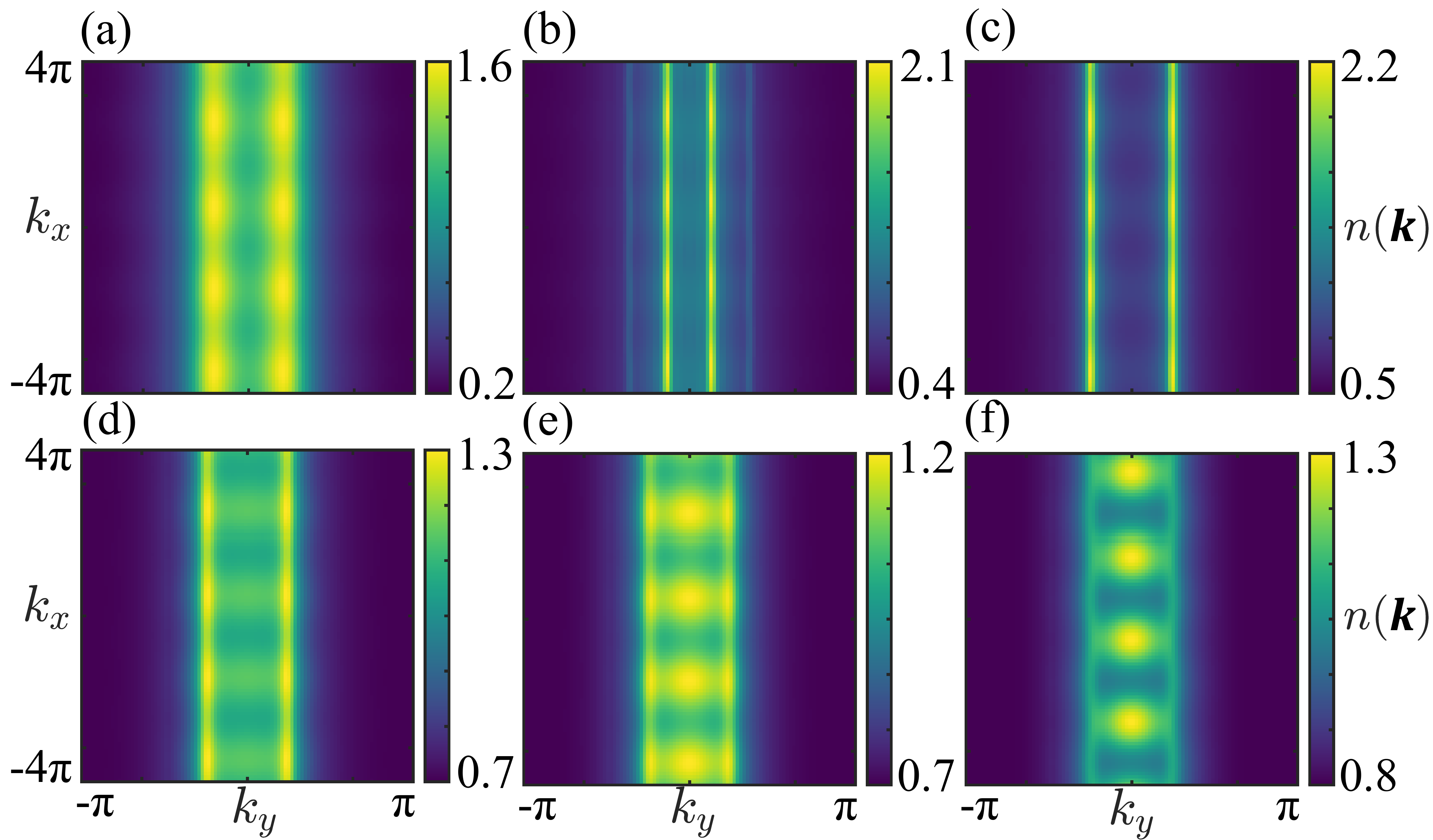}}
	\caption{Momentum profiles $n(\mathbf{k})$ for different phases when $\phi/\phi_0=0.25$ and $J/U=0.4$, including the (a) incompressible phase at $\mu/U=0$ with Meissner currents, (b) SF phase at $\mu/U=0.2$ with strong vortex-Meissner currents, (c) SF phase at $\mu/U=0.5$ with Meissner currents and several incompressible phases at (d) $\mu/U=0.77$, (e) $\mu/U=0.84$ and (f) $\mu/U=0.9$, which contain strong vortex-Meissner currents. SF phases with Meissner order contain two degenerate momentum peaks, which extend across multiple Brillouin zones. At larger $\mu/U$, a range of incompressible phases can also form, which possess finite currents.}
	\label{figure_nk2_H}
\end{figure}

By tuning the magnetic flux, it is possible to change the distributions of current within the system, as shown in Fig.~\ref{figure_phi_H}(b) for $\phi / \phi_0 = 0.25$ and $J/U=0.4$. Note, this value of $J/U$ is chosen such that when $\mu/U \approx 1$, the currents will become zero, i.e. a trivial pattern. We also plot momentum profiles again in Fig.~\ref{figure_nk2_H} for this case. The Meissner current $j_M$ is now finite for the considered flux, with an incompressible phase for $\mu/U<0.1$. By inspecting the locality measures in Fig.~\ref{figure_phi_H}(d), we observe that $j_L\gg0$ and $j_E \rightarrow 1$, implying the absence of bulk currents and formation of Meissner order. Due to this, the momentum profile in Fig.~\ref{figure_nk2_H}(a) possesses two distinct and symmetric momentum peaks \cite{PhysRevA.83.055602}, which are extended along $k_x$ at $k_y \approx \pm 3\pi/16$. For $0.1 \le \mu/U \le 0.4$, the vertical currents $j_V$ in the bulk will increase, marking the onset of a strong vortex-Meissner pattern, since horizontal currents are aligned on each row, i.e. $|j_M| = j_H$. The corresponding $n(\mathbf{k})$ in Fig.~\ref{figure_nk2_H}(b) is then more localised as a result of this. When $0.4 \le \mu/U \le 0.65$, the locality measures will again fluctuate near $1$, revealing the formation of Meissner currents, with sharp momentum peaks in Fig.~\ref{figure_nk2_H}(c). Finally, for $\mu / U > 0.65$, we enter an interesting domain in which many distinct incompressible phases appear. By inspecting Fig.~\ref{figure_nk2_H}(f), each incompressible phase corresponds to the different plateaus in $\bar{\rho}$, which all have a strong vortex-Meissner current pattern. This is also reflected in the momentum profiles from Figs.~\ref{figure_nk2_H}(d-f), which combine characteristics from both vortex and Meissner domains.

\begin{figure}[t]
	\centering
	\makebox[0pt]{\includegraphics[width=0.99\linewidth]{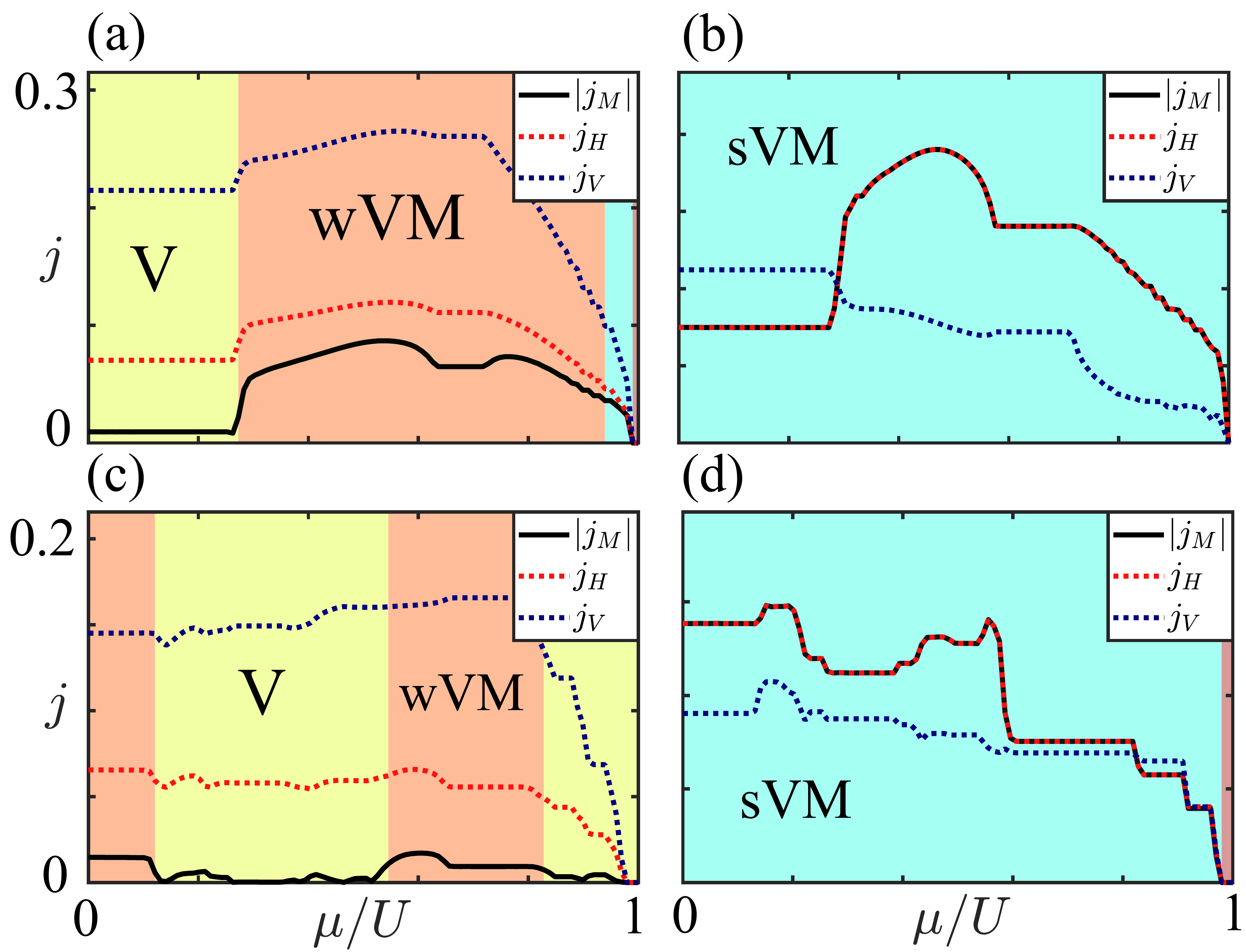}}
	\caption{Plots of current order parameters over a range of $\mu/U$ for (a,b) superlattice and (c,d) quasiperiodic ladders when $a/b=2$. We consider fixed (a,c) $J/U=0.29$, $\phi/\phi_0=0.5$ and (b,d) $J/U=0.28$, $\phi/\phi_0=0.25$, with current domains coloured according to the definitions in Table~\ref{table_phases}. Due to the non-uniform $X(m)$, significant differences can be observed with the currents, with the vortex-Meissner patterns now dominant.}
	\label{figure_phi_I}
\end{figure}

\subsection{Inhomogeneous Ladders ($a/b=2$)}
Here, we will now study the influence of inhomogeneous $X(m)$ throughout the lattice, with the distributions of bond length defined from the two cases outlined in Sec.~\ref{sc_ladders}, i.e. the superlattice and quasiperiodic ladders. We plot the behaviour of currents for these lattices in Fig.~\ref{figure_phi_I} over a range of $\mu/U$ and $J/U$, for $a/b=2$ and fixed $\phi/\phi_0$. The superlattice distribution is first considered in Figs.~\ref{figure_phi_I}(a,b), which shows that extended vortex-Meissner domains appear with far greater frequency. As a consequence, the Meissner distribution for the superlattice is no longer stable in these regions, being replaced by either strong or weak vortex-Meissner currents. By considering a quasiperiodic lattice in Figs.~\ref{figure_phi_I}(c,d), similar properties are observed, with the strong vortex-Meissner pattern dominating at $\phi/\phi_0=0.25$. For $\phi/\phi_0=0.5$, however, Meissner currents are suppressed on the quasiperiodic ladder, leading to the more frequent appearance of vortex distributions.

\begin{figure}[t]
	\centering
	\makebox[0pt]{\includegraphics[width=0.99\linewidth]{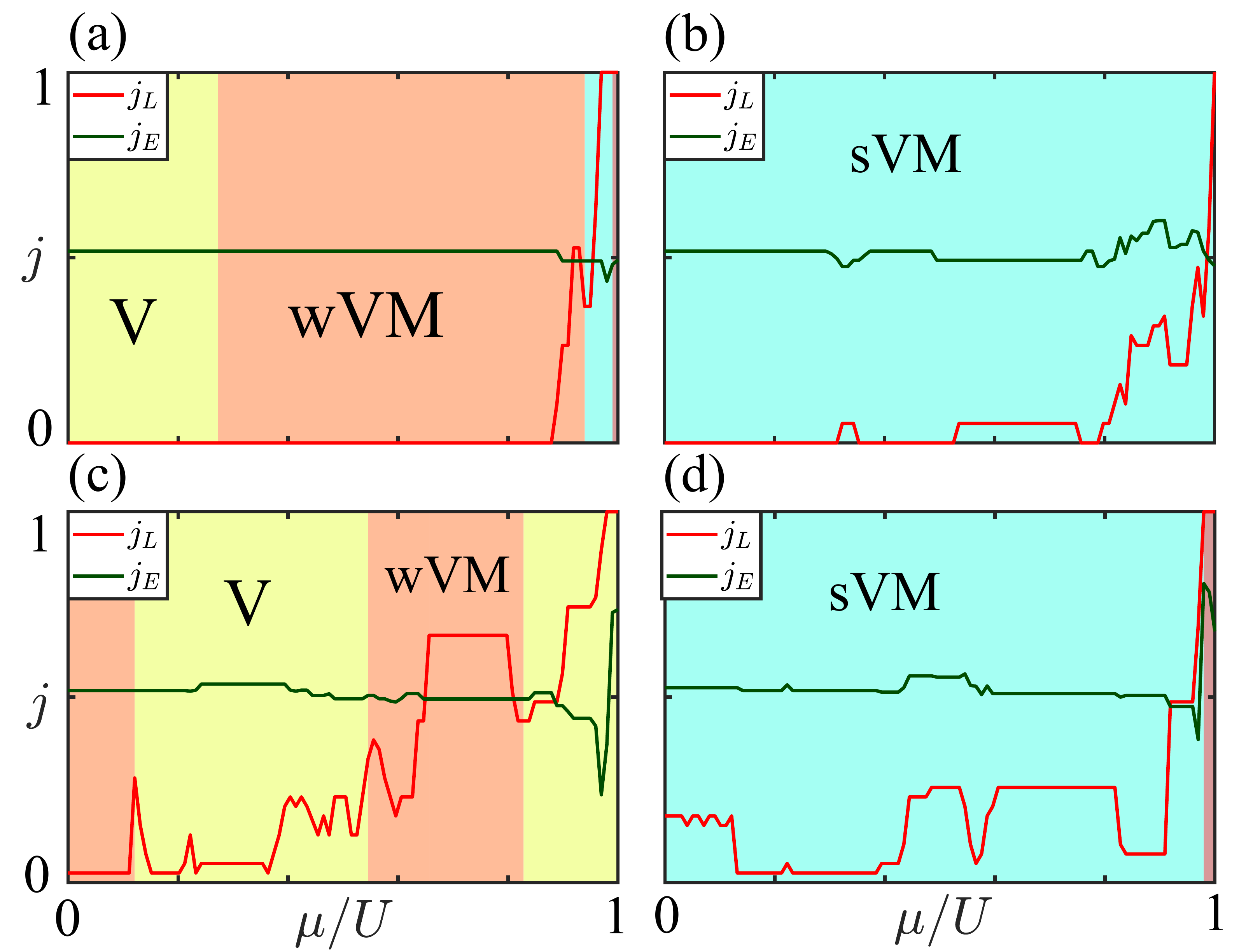}}
	\caption{Current locality measures $j_L$ and $j_E$ for (a,b) superlattice and (c,d) quasiperiodic ladders, using the same parameters and boundaries as Fig.~\ref{figure_phi_I}. Current domains are coloured according to the definitions in Table~\ref{table_phases}. The superlattice ladder has the majority of phases extended across the system, indicated by $j_E\approx0.5$ and $j_L\approx0$. On the other hand, for the quasiperiodic ladder, fluctuations in $j_L$ are more significant, marking the onset of localised phases.}
	\label{figure_phiLoc_I}
\end{figure}

To better understand the differences between homogeneous and inhomogeneous ladders, we also consider the locality measures in Fig.~\ref{figure_phiLoc_I}. For the superlattice in Figs.~\ref{figure_phiLoc_I}(a,b), we find that most currents are extended across the lattice, with $j_E\approx0.5$ and $j_L\approx 0$. Fluctuations will occur near $\mu/U\rightarrow 1$, which then marks the onset of more localised current distributions. On the other hand, the quasiperiodic ladder in Figs.~\ref{figure_phiLoc_I}(c,d) contains a higher degree of localised currents across all $\mu/U$. This is a consequence of the inhomogeneous $X(m)$ distribution, which induces a form of preferential localisation into the system. While the superlattice lacks this kind of localisation, the inhomogeneous Peierls phases and tunnelling rates can still allow for changes to the current patterns and support of vortex-Meissner domains.

\begin{figure}[t]
	\centering
	\makebox[0pt]{\includegraphics[width=0.99\linewidth]{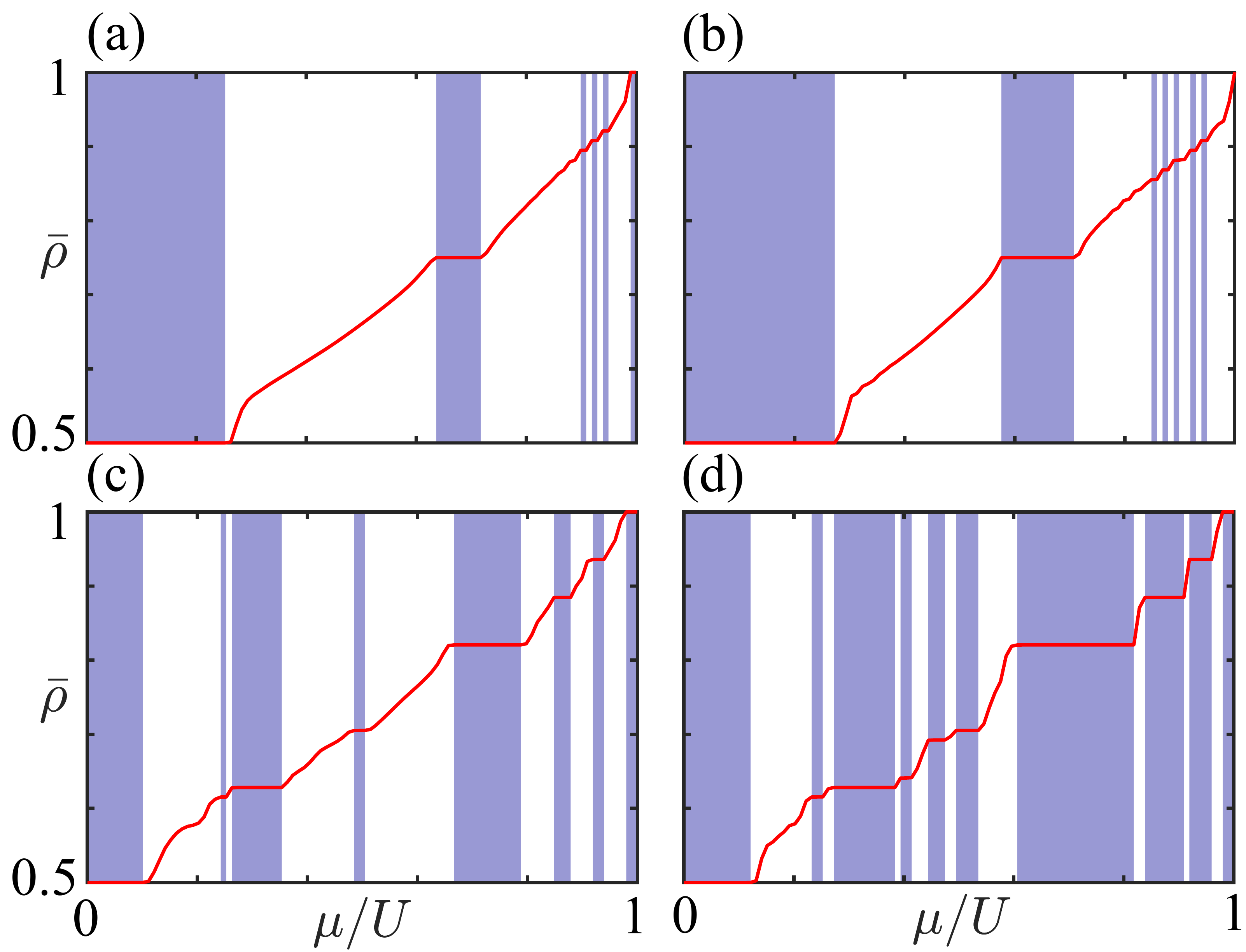}}
	\caption{Average density $\bar{\rho}$ for (a,b) superlattice and (c,d) quasiperiodic ladders, using the same parameters and boundaries as Fig.~\ref{figure_phi_I}. The shaded areas highlight the incompressible phases, i.e. the plateaus in $\bar{\rho}$. In both cases, inhomogeneous ladders can stabilise extended incompressible domains. These properties are further exaggerated for the quasiperiodic ladder.}
	\label{figure_phiRho_I}
\end{figure}

Finally, we will consider the extent of incompressible phases, which can be found by plotting $\bar{\rho}$ in Fig.~\ref{figure_phiRho_I}. We observe that the superlattice in Figs.~\ref{figure_phiRho_I}(a,b) can contain extended, incompressible domains at far greater frequency when compared against the homogeneous system in Fig.~\ref{figure_phi_H}. This includes an incompressible phase at half-filling for $\mu/U < 0.3$ and smaller domains for $\mu/U > 0.6$. For a quasiperiodic lattice, the incompressible phases are found to be far more dominant, as shown in in Figs.~\ref{figure_phiRho_I}(c,d). For certain $\phi/\phi_0$, it is also possible for the incompressible phases to appear more commonly than that of the SF.

\begin{figure}[t]
	\centering
	\makebox[0pt]{\includegraphics[width=0.99\linewidth]{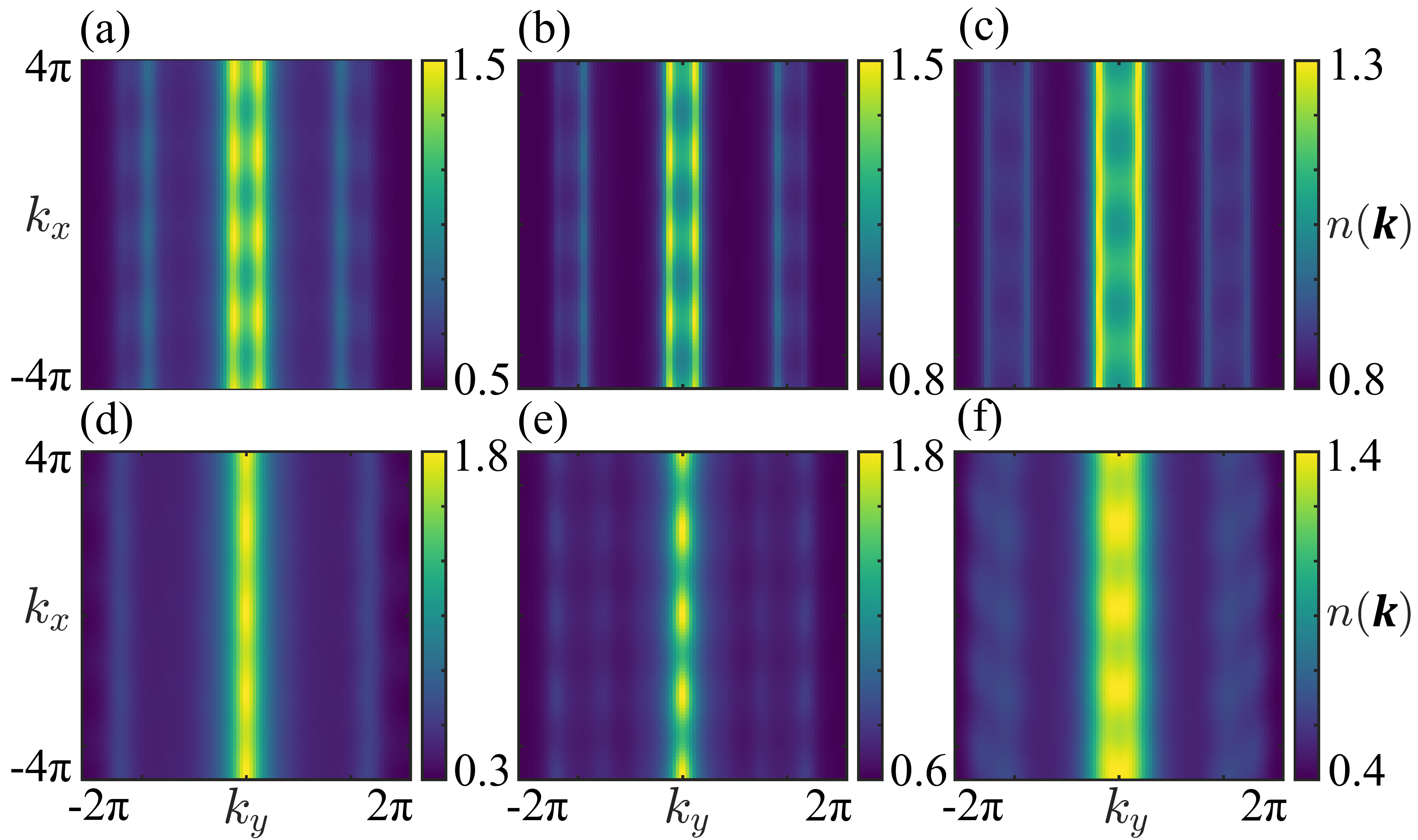}}
	\caption{Momentum profiles $n(\mathbf{k})$ of incompressible phases for (a-c) superlattice and (d-f) quasiperiodic ladders. The magnetic flux and tunnelling are fixed to (a,b,d,e) $\phi/\phi_0=0.5$ and $J/U=0.29$, and (c,f) $\phi/\phi_0=0.25$ and $J/U=0.28$, with chemical potentials of (a) $\mu/U=0.68$, (b) $\mu/U=0.92$, (c) $\mu/U=0.92$, (d) $\mu/U=0.32$, (e) $\mu/U=0.75$ and (f) $\mu/U=0.52$. Fluctuating bond lengths $a$ and $b$ produces $n(\mathbf{k})$ with mixed vortex and Meissner characteristics.}
	\label{figure_nk_I}
\end{figure}

To better visualise the incompressible phases for the inhomogeneous systems, we also plot several momentum profiles in Fig.~\ref{figure_nk_I}. For the superlattice phases in Figs.~\ref{figure_nk_I}(a-c), two symmetric momentum peaks will again appear around $k_x=0$, which vary in structure and extent for different $\mu/U$ and $\phi/\phi_0$. When $k_y \rightarrow \pm 2\pi$, we also observe the formation of separate stripe features due to the presence of inhomogeneous $X(m)$. The quasiperiodic states in Figs.~\ref{figure_nk_I}(d-f) share similar properties, although the degenerate momentum peak across $k_x$ tends to collapse towards $k_x=0$. Note, while we have considered momentum profiles of incompressible phases, the compressible domains are very similar in structure; but with further localisation of momentum peaks.


For crystalline ladders, there will be no specific regions for particles to preferentially localise towards over another, generally leading to extended structures of currents. Alternatively, for a quasiperiodic ladder, fluctuations in $X(m)$ allows for localisation throughout the lattice, with particular arrangements of current that are not be repeated.

By looking at the momentum profiles, several distinctions can also be made between different current patterns. Any phase with finite currents will be shifted away from $\mathbf{k}=(0,0)$. For vortex currents, these shifts will only be along the $k_x$ direction. Meissner currents, however, will contain two degenerate momenta peaks across $k_y$, with the shift along $k_x$ varying. The vortex-Meissner $n(\mathbf{k})$ can contain distinct properties from both vortex and Meissner distributions of $n(\mathbf{k})$. First, the strong vortex-Meissner pattern has pronounced momentum peaks, like that of the Meissner distribution, but with an envelope function that corresponds to vortex currents. The weak vortex-Meissner pattern then contains less pronounced momentum peaks, which also have a smaller relative separation. Both the superlattice and quasiperiodic ladder can form similar kinds of phases, with separate features forming along $k_y$ due to the non-uniform bond lengths. For the quasiperiodic ladders, the separation between momentum peaks usually vanishes, with larger fluctuations in the stripe-like structure.

\begin{figure*}[t!]
	\centering
	\makebox[0pt]{\includegraphics[width=0.99\linewidth]{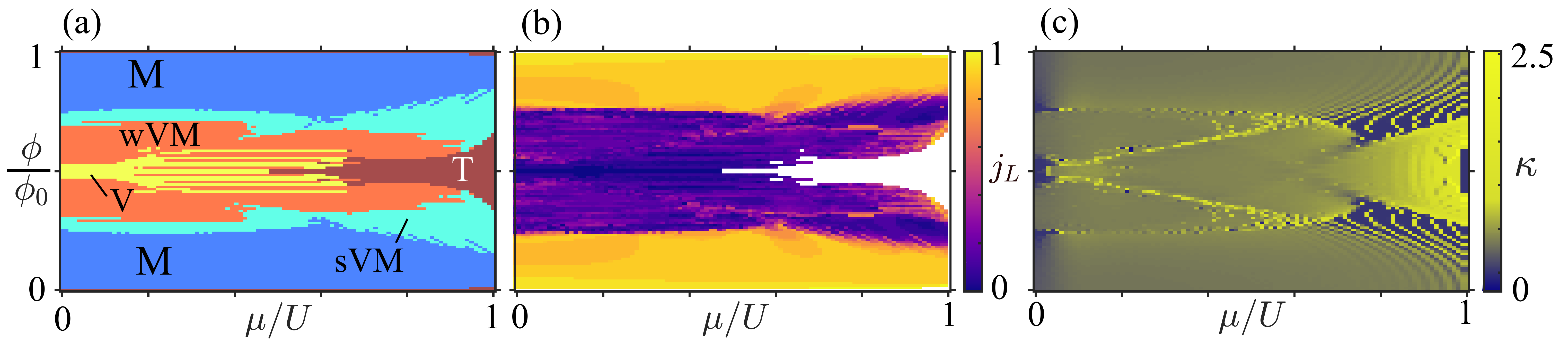}}
	\caption{Phase diagram of the homogeneous ladder for $a/b=1$ and $J/U=0.44$, showing the (a) current patterns, (b) locality $j_L$ and (c) compressibility $\kappa$. The current patterns are defined according to the definitions in Table~\ref{table_phases}. We observe a clear transition of vortex to Meissner currents, with the vortex-Meissner distributions separating their domains. For larger $\mu/U$, sets of incompressible phases can frequently appear.}
	\label{figure_phiMu_H}
\end{figure*}

\begin{figure*}[t!]
	\centering
	\makebox[0pt]{\includegraphics[width=0.99\linewidth]{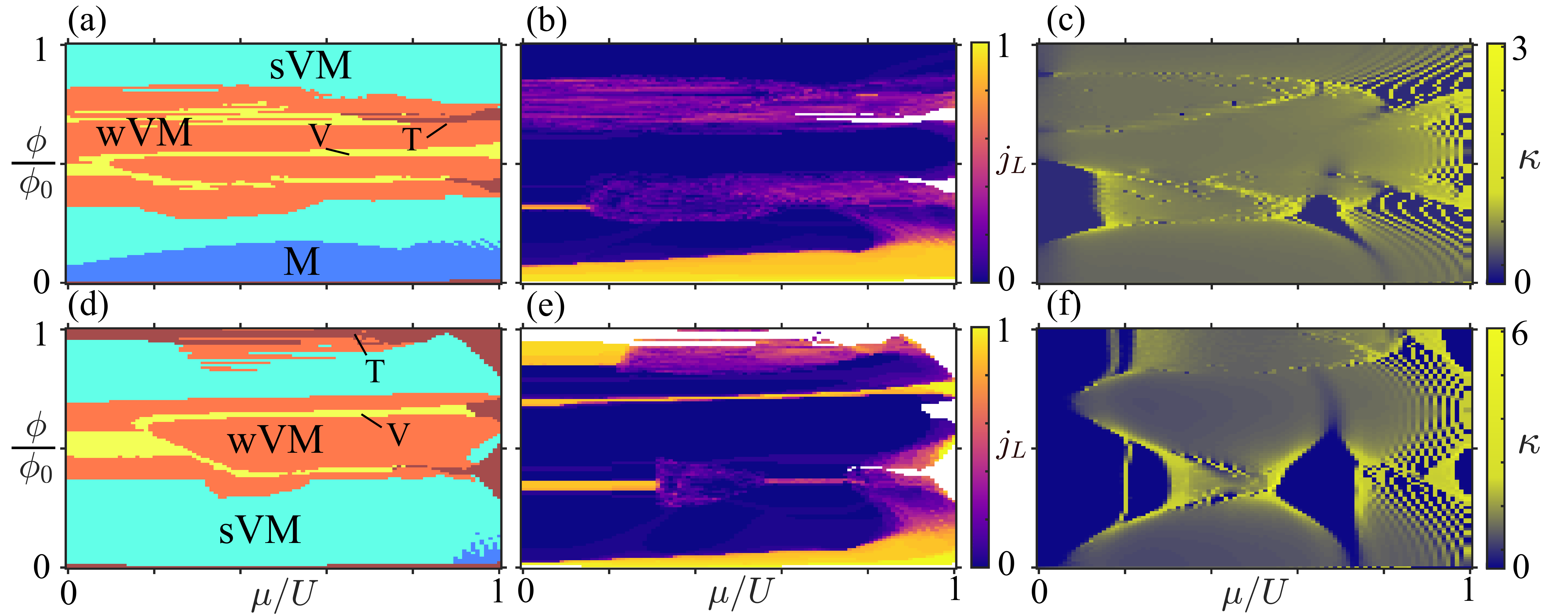}}
	\caption{Phase diagrams of the superlattice ladders for (a-c) $a/b = 1.25$ and $J/U=0.41$, and (d-f) $a/b = 2$ and $J/U=0.29$, showing (a,d) current patterns, (b,e) locality $j_L$ and (c,f) compressibility $\kappa$. The current patterns are defined according to the definitions in Table~\ref{table_phases}. The vortex-Meissner distributions are seen to dominate many of the phase regions when compared against the homogeneous ladder, with certain incompressible regions also growing in extent.}
	\label{figure_phiMu_S}
\end{figure*}

\begin{figure*}[t!]
	\centering
	\makebox[0pt]{\includegraphics[width=0.99\linewidth]{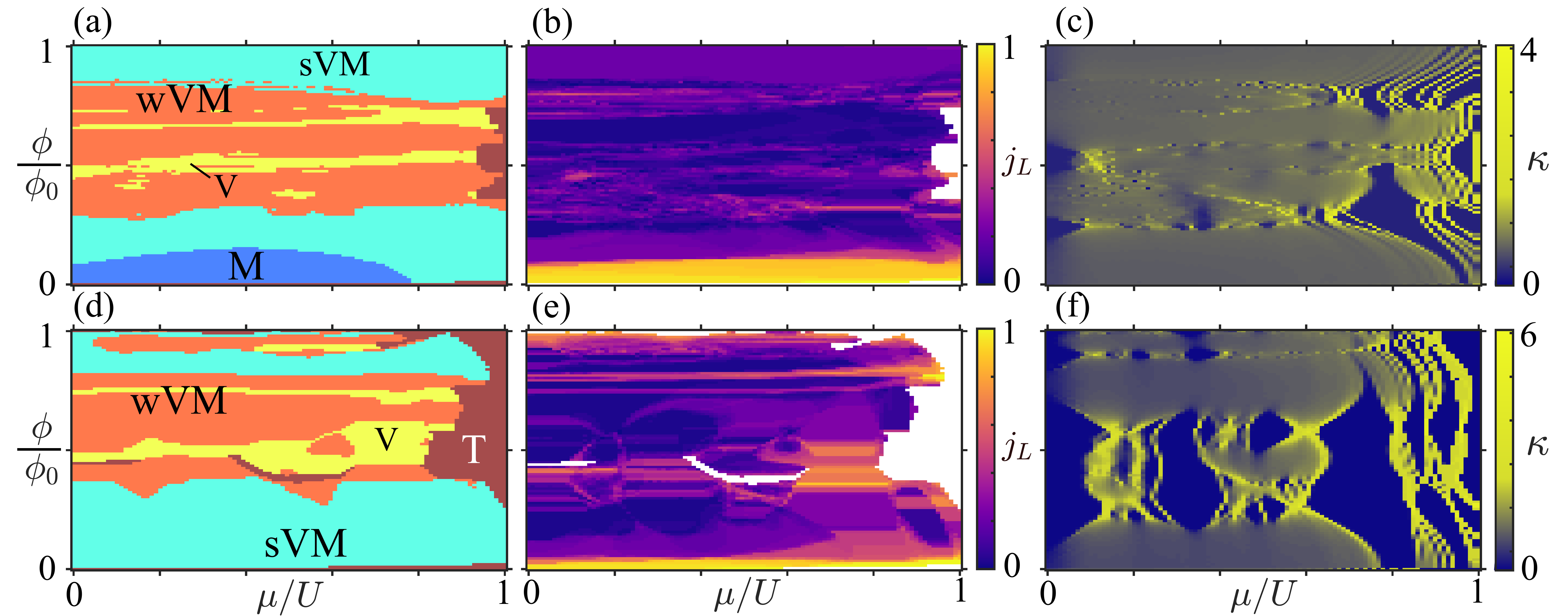}}
	\caption{Phase diagrams of the quasiperiodic ladders for (a-c) $a/b = 1.25$ and $J/U=0.41$, and (d-f) $a/b = 2$ and $J/U=0.29$, showing (a,d) current patterns and (b,e) locality $j_L$ and (c,f) compressibility $\kappa$. These regions are defined according to the definitions in Table~\ref{table_phases}. Here, the incompressible phases and vortex-Meissner domains again grow in size and extent. The locality $j_L$ is also larger when compared to the superlattice and homogeneous ladders, which indicates the presence of localised currents.}
	\label{figure_phiMu_Q}
\end{figure*}

\section{Magnetic Phase Diagrams} \label{sc_mDiag}
We will now plot full phase diagrams for the different ladders as a function of $\mu/U$ and $\phi/\phi_0$, labelling currents according to the definitions of Table~\ref{table_phases}, as before. The compressibility $\kappa$ will be plotted, which can be used to visualise whether or not the system is in a SF-like or incompressible phase. This will be particularly important for the superlattice and quasiperiodic ladders, in which incompressible domains were prominent. Furthermore, inhomogeneous ladders can localise the current distributions, i.e. $j_L \gg 0$. If $\kappa$ is finite when $j_L \gg 0$, we therefore have a BG-like phase, rather than a macroscopic SF.

\subsection{Homogeneous Ladder}
For the first set of results, we will consider the homogeneous $a/b=1$ ladder in Fig.~\ref{figure_phiMu_H}, for fixed $J/U=0.44$. This particular choice of $J/U$ is chosen such that when $\mu/U \rightarrow 1$, the phase converges towards the MI phase with zero (trivial) currents. In other words, the range of $\mu/U$ we consider will characterise all finite current patterns. Phase regions over $-1 \le \mu/U \le 0$ will also be mirrored due to the presence of particle-hole symmetry in the hard-core, Bose-Hubbard model. In Fig.~\ref{figure_phiMu_H}(a), we colour different regions according to the current distributions in Table~\ref{table_phases}. For $\phi/\phi_0 \approx 0.7$, there is a clear transition from the vortex to Meissner patterns of current, as has been observed in prior studies \cite{PhysRevB.91.140406,Orignac_2016}. The current regions are also symmetric about $\phi/\phi_0=0.5$, in a similar manner to the single-particle Hofstadter butterfly \cite{PhysRevB.14.2239}. If $\phi/\phi_0$ approaches an integer, the trivial pattern of current can be stabilised. This also occurs for a range of $\phi/\phi_0 \approx 0.5$, near $\mu/U \rightarrow 1$. Vortex-Meissner domains will also appear throughout the considered regions for smaller ranges of $\phi/\phi_0$, and will usually separate vortex and Meissner domains. The current locality $j_L$ is also plotted in Fig.~\ref{figure_phiMu_H}(b), which further illustrates the clear transition between vortex (extended) and Meissner (localised) currents. In other words, a $j_L \approx 1$ denotes an absence of bulk currents across vertical bonds, i.e. Meissner order. For $j_L \approx 0$, currents are fully extended, indicating variations of the vortex pattern. Fluctuations will of course arise due to the different patterns of vortex currents and changes in the magnetic unit cell. However, all phases found for the homogeneous ladder will retain an extended structure of currents throughout the lattice. Finally, in Fig.~\ref{figure_phiMu_H}(c), we also plot the compressibility $\kappa$, which can be used to characterise the transition between incompressible and SF phases. Here, we can see that the SF phase is usually dominant. However, we note the presence of several incompressible domains around $\mu/U=0$, which contain vortex or Meissner current patterns. Furthermore, when $0.7 \le \mu/U \le 1.0$, we observe the formation of a large set of incompressible phases with vortex-Meissner patterns of current.

\subsection{Superlattice Ladder}
Here, we now consider phase diagrams for the superlattice ladders in Fig.~\ref{figure_phiMu_S}, for different values of $a/b$ and $J/U$. If we first consider a small variation to $a/b=1.25$, as per Figs.~\ref{figure_phiMu_S}(a-c), there are several similarities to the homogeneous ladder in terms of the general location of current distributions and incompressible domains. However, several key differences are also observed due to the inhomogeneous $X(m)$ distribution. This includes the two larger incompressible phases around $\phi/\phi_0=0.3$ at $\mu/U=0.1$ and $\mu/U=0.67$, and a restriction of the Meissner current patterns to a smaller range of $\phi/\phi_0$ below $0.2$. Furthermore, the weak and strong vortex-Meissner crossover domains will then occupy larger regions of the phase diagram, which is a consequence of inhomogeneous tunnelling across the different rectangular tiles. If we inspect the current locality $j_L$ in Figs.~\ref{figure_phiMu_S}(b), it can be seen that $j_L\approx0$ in most domains, which indicates the presence of extended currents across the lattice. When $a/b$ is further deviated to $2$ in Figs.~\ref{figure_phiMu_S}(d-f), these differences will be exaggerated further, with incompressible domains growing in size and overall extent. The Meissner distribution is also restricted to a very small range of $\phi/\phi_0\approx0$ in Fig.~\ref{figure_phiMu_S}(d), being replaced primarily by strong vortex-Meissner domains. We also observe small values of $j_L$ within Figs.~\ref{figure_phiMu_S}(e), which again reveals the extended nature of superlattice currents.

\subsection{Quasiperiodic Ladder}
In this final section, we will consider the phase diagrams of the quasiperiodic ladders in Fig.~\ref{figure_phiMu_Q}, for different values of $a/b$ and $J/U$. As before, we initially consider a small variation of $a/b=1.25$ in Figs.~\ref{figure_phiMu_Q}(a-c), which has similar features to what was observed with the superlattice ladder in Figs.~\ref{figure_phiMu_S}(a-c), including the growth of incompressible phases and destabilisation of the Meissner domains. However, the current locality in Figs.~\ref{figure_phiMu_S}(b) tends to fluctuate around non-zero values across larger domains, showing that localised currents are forming on the quasiperiodic ladder. The differences between both homogeneous and superlattice ladders becomes significant when a larger deviation of $a/b=2$ is used in Figs.~\ref{figure_phiMu_Q}(d-f). For these cases, incompressible phases and strong/weak vortex-Meissner domains become far more dominant across most $\mu/U$ and $\phi/\phi_0$, as expected from Fig.~\ref{figure_phiMu_S}, with the Meissner distribution being absent. Furthermore, $j_L$ in Figs.~\ref{figure_phiMu_Q}(e) contains a greater degree of locality away from a value of $0$, which indicates the formation of intriguing, bulk localised currents. There are also several regions in which both $j_L$ and $\kappa$ are finite, which implies the formation of BG-like phases on the ladder. This is to be expected due to the absence of short-range order for the quasiperiodic ladder, i.e. different structures of current can not be repeated, and are hence localised within the bulk.

\section{Conclusions} \label{sc_conc}
In summary, we have shown that different kinds of ladder systems can possess exotic incompressible phases and localisation properties. If the ladders are homogeneous or arranged as a superlattice, we observe extended and periodic structures of current across the system, with vortex and Meissner patterns dominant. On the other hand, by considering ladders with greater deviations in the bond lengths $a/b$, it is possible to induce dramatic shifts in the overall phase regions. For the quasiperiodic and superlattice ladders, we have seen that the weak or strong vortex-Meissner domains can become much more prominent. Furthermore, incompressible phases can occupy larger regions of the parameter space. These properties are a direct consequence of the fluctuating bond lengths and corresponding encircling phases $\Theta$, which allows for preferential localisation within the bulk, towards the smaller bond lengths. This has been made particularly clear for the case of quasiperiodic ladders, in which there is a high degree of current locality for both compressible and incompressible phases.


\begin{acknowledgments}
	D.J. acknowledges support from EPSRC CM-CDT Grant No. EP/L015110/1. C.W.D. acknowledges support by the EPSRC Programme Grant DesOEQ (EP/T001062/1).
\end{acknowledgments}

%


\end{document}